\documentclass[prl,twocolumn,showpacs]{revtex4}
\usepackage{graphicx,epsfig}
\usepackage{times,amsmath,amssymb,latexsym}
\usepackage{rotating}
\usepackage{stmaryrd}
\usepackage{isolatin1}

\newcommand{\eqn}[1]{\begin{eqnarray} #1 \end{eqnarray}}
\newcommand{\abs}[1]{\left| #1 \right|}

\newcommand{\equals}{&\!\!\!=\!\!\!&}

\begin{document}

\title{Fractional Quantum Hall States in Ultracold Rapidly Rotating Dipolar
Fermi Gases.}
\author{M.A. Baranov$^{1,2}$, Klaus Osterloh$^{1}$, and M. Lewenstein$^{1}$}
\affiliation{(1) Institute for Theoretical Physics, University of Hannover,
Appelstr. 2, 30167 Hannover\\
(2) RRC "Kurchatov Institute", Kurchatov sq. 1, 123182 Moscow, Russia}

\begin{abstract}
We demonstrate the experimental feasibility of incompressible fractional
quantum Hall-like states in ultra-cold two dimensional rapidly rotating
dipolar Fermi gases. In particular, we argue that the state of the system at
filling fraction $\nu =1/3$ is well-described by the Laughlin wave function
and find a substantial energy gap in the quasiparticle excitation spectrum.
Dipolar gases, therefore, appear as natural candidates of systems that allow
to realize these very interesting highly correlated states in future
experiments.
\end{abstract}

\date{\today }
\maketitle




During the recent years, cold atom systems with strongly pronounced
interparticle correlations have become a subject of intensive studies, both
theoretically and experimentally. There are several ways to increase the
role of interparticle interactions in gaseous trapped systems and to reach
the strongly correlated regime. One of the possibilities is to employ an
optical lattice where the tunneling strength between sites is smaller than
the Hubbard-like on-site interaction \cite{Jaksch:1998aa}. This approach has
led to a spectacular experimental observation of the Mott-Hubbard transition
in atomic lattice Bose gases \cite{Greiner:2002aa} and is nowadays a main
tool to create strongly correlated systems. Another way to enhance the
effects of interparticle interactions is to use a quasi-2D rotating harmonic
trap \cite{Dalibard:2000aa,Cornell:2001aa,Dalibard:2004aa,Cornell:2004aa}.
When the rotational frequency approaches the trap frequency, i.e. in the
limit of critical rotation, the single particle energy spectrum becomes
highly degenerate, and hence, the role of interparticle interactions becomes
dominant. The Hamiltonian of the system in the rotating frame of reference
is formally equivalent to the one of charged particles moving in a constant
perpendicular magnetic field. This opens a remarkable possibility to
establish a link with physics of the quantum Hall effect and to realize a
large variety of strongly correlated states proposed in the context of the
fractional quantum Hall effect (FQHE) \cite{Prange:1987aa}, in a completely
different experimental setup. Recently,
the idea of composite bosons -- bound states
of vortices and bosonic atoms -- has been successfully used to describe the
ground state of a rotating Bose-Einstein condensate in a parabolic trap in
the regime of large coherence length \cite{Wilkin:2000aa,Wilkin:2001aa}. In
Ref. \cite{Cirac:2001aa}, a method of creating, manipulating and detecting
anyonic quasi-particle excitations for fractional quantum Hall bosons at
filling fraction $\nu =1/2$ in rotating Bose-Einstein condensates has
been proposed. However, it was found that because of the short-range
character of interparticle interactions, fractional quantum Hall 
states are only
feasible for a small number of particles. This is due to the fact that
Laughlin-like states do not play any specific role, when the interaction is
short-ranged. Indeed, the Jastrow prefactor in the corresponding wave
functions, $\prod_{i<j}(z_{i}-z_{j})^{p}$, where $z_{j}=x_{j}+\mathrm{i}y_{j}
$ is the coordinate of the $j$-th particle, and $p$ is an integer (even for
bosons and odd for fermions), makes the effects of a short-range interaction
irrelevant. As a consequence, excitations are gapless and the states
themselves are compressible. This contrasts to the case of electrons where
the Coulomb interaction favors fractional quantum Hall phases by lifting the
degeneracy of the ground state by providing a gap for single-particle
excitations \cite{Prange:1987aa}. It should be noted that in some cases
(when more than one Landau level is occupied in the composite particle
description of the fractional quantum Hall effect \cite{Jain}) the situation
can be improved \cite{Jolicoeur:2004aa} by using the recently observed
Feshbach resonance in the $p$-wave channel \cite{Bohn}. This resonance,
however, is accompanied by dramatic losses that make its experimental
application questionable.

%
%
In this letter, we demonstrate that rotating quasi-2D gaseous systems with
dipole-dipole interactions could provide all necessary ingredients for the
observation of fractional quantum Hall-like states. In particular, the
dipole-dipole interaction favors fractional quantum Hall phases by creating
a substantial gap in the single-particle excitation spectrum and makes them
incompressible. We demonstrate this for the case of a quasihole excitation
in the most famous Laughlin state at filling $\nu =1/3$ in a homogeneous
quasi-2D dipolar rotating Fermi gas with dipolar moments polarized
perpendicular to the plane of motion. Furthermore, we discuss the
possibility of providing the rotating reference frame with a quenched
disorder that ensures the robust creation and observation of fractional
quantum Hall states in experiments with trapped gases.

%
%
We consider a system of $N$ dipolar fermions rotating in an axially
symmetric harmonic trapping potential with a strong confinement along the
axis of rotation, the $z$-axis. With respect to the latter, the dipoles are
assumed to be aligned. Various ways of experimental realizations of
ultracold dipolar gases are discussed in the review \cite{nobel}. Assuming
that the temperature $T$ and the chemical potential $\mu $ are much smaller
than the axial confinement, $T,\,\mu \ll \omega _{z}$, the gas is
effectively two-dimensional, and the Hamiltonian of the system in the
rotating reference frame reads 
\begin{equation}
{\mathcal{H}}\!\!=\!\!\sum_{j=1}^{N}\Big(\!\frac{p_{j}^{2}}{2m}+\frac{m}{2}%
\omega _{0}^{2}r_{j}^{2}-\omega L_{jz}\!\Big)+\sum_{j<k}^{N}\frac{d^{2}}{%
\left\vert \mathbf{r}_{j}-\mathbf{r}_{k}\right\vert ^{3}}\,.
\label{Hamiltonian}
\end{equation}%
Here $\omega _{0}\ll \omega _{z}$ is the radial trap frequency, $\omega $ is
the rotation frequency, $m$ is the mass of the particles, $d$ their dipolar
moment, and $L_{jz}$ is the projection of the angular momentum with respect
to the $z$-axis of the $j$-th particle located at $\mathbf{r}_{j}=x_{j}%
\mathbf{e}_{x}+y_{j}\mathbf{e}_{y}$. The above Hamiltonian can be
conveniently rewritten in the form 
\begin{equation}
{\mathcal{H}}\!\!=\!\!\!\sum_{j=1}^{N}\!\bigg[\,\underset{{\mathcal{H}}_{%
\mathrm{Landau}}}{\underbrace{\frac{1}{2m}\left( \mathbf{p}_{j}-m\omega _{0}%
\mathbf{e}_{z}\times \mathbf{r}_{j}\right) ^{2}\,}}\!\!+\underset{{\mathcal{H%
}}_{\Delta }}{\underbrace{(\omega _{0}-\omega )L_{jz}}}\bigg]\!+V_{\mathrm{d}%
}\,,  \label{modifHamilton}
\end{equation}%
where ${\mathcal{H}}_{\mathrm{Landau}}$ is formally equivalent to the {Landau%
} Hamiltonian of particles with mass $m$ and charge $e$ moving in a constant
perpendicular magnetic field with the vector potential $\mathbf{A}=(cm\omega
_{0}/e)\mathbf{e}_{z}\times \mathbf{r}$, $V_{\mathrm{d}}$ is the
dipole-dipole interaction (the last term in Eq.~(\ref{Hamiltonian})), and ${%
\mathcal{H}}_{\Delta }$ describes the shift of single-particle energy levels
as a function of their angular momentum and the difference of the
frequencies $\Delta \omega =\omega _{0}-\omega $.

In the limit of critical rotation $\omega \rightarrow \omega _{0}$, one has $%
{\mathcal{H}}_{\Delta }\!\!\ll \!\!\{{\mathcal{H}}_{\mathrm{Landau}},\,V_{%
\mathrm{d}}\}$, and the {Hamiltonian} \eqref{Hamiltonian} describes the
motion of dipolar particles in a constant perpendicular magnetic field with
cyclotron frequency $\omega _{\mathrm{c}}=2\omega _{0}$ \cite{Wilkin:2000aa}%
. The spectrum of ${\mathcal{H}}_{\mathrm{Landau}}$ is well-known and
consists of equidistantly spaced {Landau} levels with energies $\varepsilon
_{n}=\hbar \omega _{c}(n+1/2)$. Each of these levels is highly degenerate
and contains $N_{\mathrm{LL}}=1/2\pi l_{0}^{2}$ states per unit area, where $%
l_{0}=\sqrt{\hbar /m\omega _{c}}$ is the magnetic length. For a given
fermionic surface density $n$ one can introduce the filling factor $\nu
=2\pi l_{0}^{2}n$ that denotes the fraction of occupied Landau levels. Note
that under the condition of critical rotation, the density of the trapped
gas is uniform except at the boundary provided by an external confining
potential.

For filling fractions $\nu \leq 1$, particles solely occupy the lowest
Landau level and the corresponding many-body eigenfunction of ${\mathcal{H}}%
_{\mathrm{Landau}}$ takes the form 
\begin{equation*}
\Psi (z_{j})=\mathcal{N}P[z_{1},\,\ldots ,\,z_{N}]\exp
(-\!\sum_{j=1}^{N}\left\vert z_{j}\right\vert ^{2}/4l_{0}^{2})\,,
\end{equation*}%
where $\mathcal{N}$ is the normalization factor and $P[\left\{ z_{j}\right\}
]$ is a totally antisymmetric polynomial in the coordinates $z_{j}=x_{j}+%
\mathrm{i}y_{j}$ of the particles. The corresponding eigenenergy is
independent of the specific choice of $P[\left\{ z_{j}\right\} ]$ and equals 
$N\hbar \omega _{c}/2$, where $N$ is the total number of particles. This
degeneracy is removed if the dipole-dipole interaction $V_{d}$ is
considered. In the following, we limit ourselves to a system at filling $\nu
=1/3$, where interparticle interaction effects are most pronounced.

In this case, the trial wave functions for the ground and quasi-hole excited
states can be taken in the form proposed by Laughlin \cite{Laughlin:1983aa} 
\begin{eqnarray}
\Psi _{\mathrm{L}}(\left\{ z_{j}\right\} ) &\!\!\!=\!\!\!&{\mathcal{N}}%
\prod_{k<l}^{N}(z_{k}-z_{l})^{3}\!\exp \!\bigg(-\sum_{i}^{N}\left\vert
z_{i}\right\vert ^{2}/4l_{0}^{2}\bigg)\,,  \label{psilaugh} \\
\Psi _{\mathrm{qh}}(\left\{ z_{j}\right\} \!,\,\zeta _{0}) &\!\!\!=\!\!\!&{%
\mathcal{N}}_{0}\prod_{j=1}^{N}(z_{j}-\zeta _{0})\Psi _{\mathrm{L}}\,,
\label{psiqh}
\end{eqnarray}%
where $\zeta _{0}$ is the position of the quasi-hole. The choice of these
wave functions in the case of the considered system with dipole-dipole
interactions can be justified as follows. These are exact eigenfunctions for
short-range $\delta $-like potentials and are proven to be very good trial
wave functions for the Coulomb interaction problem. Actually, as it was
shown by Haldane (see the corresponding contribution in \cite{Prange:1987aa}%
), the Laughlin states in the fractional quantum Hall effect are essentially
unique and rigid at the corresponding filling factors ($\nu =1/3$ in our
case). They are favored by strong short-range pseudopotential components
that are particularly pronounced in the case of a dipolar potential.

Another possible candidate for the ground state of our system could be a
crystalline state similar to the 2D Wigner electron crystal in a magnetic
field \cite{Wigner}. For a non-rotating dipolar Fermi gas in a 2D trap, this
state has lower energy than the gaseous state for sufficiently high
densities. The estimate of the stability region can be obtained from the
Lindemann criterion: the ratio $\gamma $ of the mean square difference of
displacements in neighbouring lattice sites to the square of the
interparticle distance, $\gamma =\left\langle \left( \mathbf{u}_{i}-\mathbf{u%
}_{i-1}\right) ^{2}\right\rangle /a^{2}$, should be less than some critical
value $\gamma _{c}$ (see, e.g., Ref. \cite{lindemann}). The results Ref. 
\cite{Lozovik} indicate that $\gamma _{c}\approx 0.07$. For zero
temperature, $\gamma $ could be estimated as $\gamma \sim \hbar
/ma^{2}\omega _{D}$, where $\omega _{D}$ is the characteristic (Debye)
frequency of the lattice phonons, $\omega _{D}^{2}\sim 36d^{2}/ma^{5}$. As a
result, the dipolar crystal \ in a non-rotating gas is stable if the
interparticle distance $a=(\pi n)^{-1/2}$ satisfies the condition $%
a<a_{d}(6\gamma _{c})^{2}\ll a_{d}$, i.e., the gas is in the strongly
correlated regime, $V_{d}\sim d^{2}/a^{3}\sim (a_{d}/a)(\hbar
^{2}/ma^{2})\gtrsim \varepsilon _{F}/(6\gamma _{c})^{2}\gg \varepsilon _{F}$.

A strong magnetic field with the cyclotron frequency larger than the Debye
frequency, $\omega _{c}>\omega _{D}$, favors the crystalline state by
modifying the vibrational spectrum of the crystal. In this case, $\gamma
\sim \hbar /ma^{2}\omega _{c}$ \cite{Wigner}, and the corresponding critical
value is $\gamma _{c}\approx 0.08$ \cite{Lozovik}. Therefore, the
crystalline state is stable if $\gamma \sim \nu /2<\gamma _{c}$. This limits
the filling factor $\nu $ to small values $\nu <1/6$. As a result, the
ground state of the system at filling factor $\nu =1/3$ is indeed
well-described by the Laughlin wave function (\ref{psilaugh}).

In order to proof the incompressibility of the state $\Psi _{\mathrm{L}}$,
we calculate the energy gap $\Delta \varepsilon _{\mathrm{qh}}$ for the
quasihole excitation. This gap can be expressed in terms of the pair
correlation functions of the ground state $g_{0}(z_{1},\,z_{2})$ and the
quasihole state $g_{\mathrm{qh}}(z_{1},\,z_{2})\!$, respectively, as

\begin{equation}
\Delta \varepsilon _{\mathrm{qh}}\!\!\!=\!\!\!\!\!\int \!\!\!\mathrm{d}%
^{2}z_{1}\mathrm{d}^{2}z_{2}V_{\mathrm{d}}(z_{1},\,z_{2})\left( g_{\mathrm{qh%
}}(z_{1},\,z_{2})\!-\!g_{0}(z_{1},\,z_{2})\right) \,.  \label{gapexpression}
\end{equation}%
For the states (\ref{psilaugh}) and (\ref{psiqh}), the functions $g_{\mathrm{%
0}}$ and $g_{\mathrm{qh}}$ have been calculated using the Monte Carlo method 
\cite{MonteCarlo}; they were approximated by {Girvin }\cite{Girvin:1984aa}
in the thermodynamic limit as 
\begin{widetext}
\vspace*{-0.7cm}
\begin{subequations}
\eqn{g_{0}(z_1,\,z_2)\equals \frac{\nu^2}{(2\pi)^2}\!\!\Big(1-{\rm
e}^{-\frac{\abs{z_1-z_2}^2}{2}}
-2\sum_{j}^{\rm \scriptscriptstyle odd}\frac{C_j}{4^jj!}\abs{z_1-z_2}^{2j}{\rm
e}^{-\frac{\abs{z_1-z_2}^2}{4}}\!\!\Big)\\
\label{gqh}
g_{\rm qh}(z_1,\,z_2)\equals\frac{\nu^2}{(2\pi)^2}\bigg[\prod_{j=1}^2
\Big(1-{\rm
e}^{-\frac{\abs{z_j}^2}{2}}\Big)
-{\rm e}^{-\frac{\abs{z_1}^2+\abs{z_2}^2}{2}}
\bigg(\Big|{\rm e}^{\frac{z_1z^\star_2}{2}}-1\Big|^2+
2\sum_{j}^{\rm \scriptscriptstyle odd}\frac{C_j}{4^jj!}\sum_{k=0}^\infty
\frac{\abs{F_{j,\,k}(z_1,\,z_2)}^2}{4^kk!}
\bigg)\bigg]\,,\\
&&F_{j,\,k}(z_1,\,z_2)=\frac{z_1z_2}{2}
\sum_{r=0}^{j}\sum_{s=0}^{k}
{j \choose r}{k \choose s}
\frac{(-1)^rz_1^{r+s}z_2^{j+k-(r+s)}}{\:\:\sqrt{(r+s+1)(j+k+1-(r+s))}\:\:}\:.}
\end{subequations}
\end{widetext}With respect to $\nu =1/3$, it was argued that an accuracy
better than $2\%$ is already achieved when only the first two coefficients $%
C_{1}\!=\!1$ and $C_{3}\!=\!-1/2$ are taken into account. In Fig.1 we plot
the difference $g_{\mathrm{qh}}\!-\!g_{0}$ for the particular choice $z_{1}=3
$ and $\zeta _{0}=0$. After substituting these expressions into Eq. (\ref%
{gapexpression}) and integrating numerically, we obtain 
\begin{equation}
\Delta \varepsilon _{\mathrm{qh}}\!\!\!=\!\!\!(0.9271\pm
0.019)\,d^{2}/l_{0}^{3}  \label{gapnumber}
\end{equation}%
for the energy gap in the spectrum of quasiholes. Naturally, a gap of the
same order of magnitude is to be expected in the spectrum of quasiparticles
(quasielectrons, in the language of the fractional quantum Hall effect),
although calculations in this case are much more difficult because , to the
best of our knowledge, the closed or even approximate expression for the
corresponding pair correlation function does not exist. The gap (\ref%
{gapnumber}) can also be written in the form 
\begin{equation*}
\Delta \varepsilon _{\mathrm{qh}}\!\!\!=\!\!\!(0.9271\pm 0.019)\hbar \omega
_{c}(a_{d}/l_{0}),
\end{equation*}%
where $a_{d}=md^{2}/\hbar ^{2}$ can be considered as a characteristic size
of the dipole interaction. For a dipole moment of the order of $0.5\mathrm{%
Debye}$, mass $m\sim 30$\textrm{\ }atomic mass units, the value of $a_{d}$
is of the order of $10^{3}\mathring{A}$, and for the trap frequency $\omega
_{0}\sim 2\pi 10^{3}\mathrm{Hz}$, one obtains the gap $\Delta \varepsilon _{%
\mathrm{qh}}\sim 30\mathrm{nK}$ and the ratio $\Delta \varepsilon _{\mathrm{%
qh}}\!\!\!/\hbar \omega _{c}<1$. This result shows (see Fig. 2) that on the
one hand, the interparticle interaction does not mix different Landau
levels, and thus the lowest Landau level approximation used in the
construction of the Laughlin wave function (\ref{psilaugh}) is reliable. On
the other hand, it guarantees that the neglected term ${\mathcal{H}}_{\Delta
}$ in the Hamiltonian (\ref{modifHamilton}), which is inevitably present in
some experimental realizations (see below), is indeed small and does not
influence the trial wave functions.

\begin{figure}[tbp]
\includegraphics[width=0.45\textwidth]{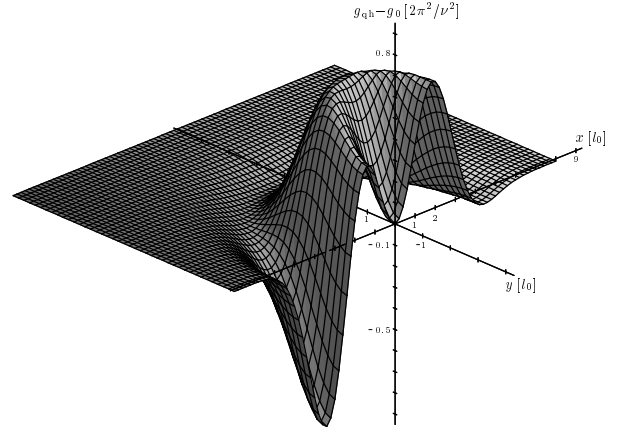}
\caption{The difference $g_{\mathrm{qh}}-g_{0}$ as a function of $z\equiv
z_{2}-z_{1}$ for $\protect\zeta _{0}=0$ and $z_{1}=3$. Both particles
strongly avoid each other, and rotational invariance is broken by the
quasi-hole.}
\end{figure}
\begin{figure}[tbp]
\includegraphics[width=0.40\textwidth]{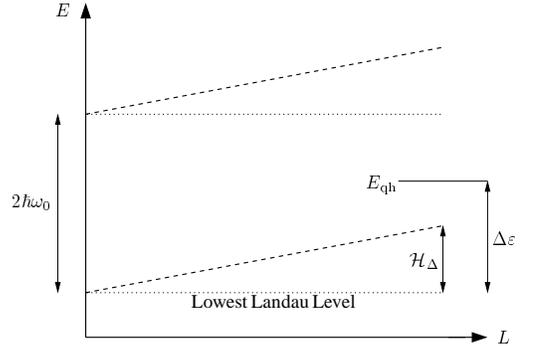}
\caption{Single-particle nergy levels of the Hamiltonian
${\mathcal{H}}={\mathcal{H}}_{\mathrm{Landau}}+{\mathcal{H}}_{\Delta }$
versus their angular momentum $L$.}
\end{figure}

Let us now discuss possible ways of experimental realization and detection
of the above states. At present, there exist two experimental methods to
create rapidly rotating gas clouds. In the JILA experiments \cite%
{Cornell:2004aa,Cornell:2001aa}, a rotating condensate in the harmonic trap
was created by evaporation of one of the spin components, and rotational
rates $\omega >99\%$ of the centrifugal limit were achieved. In this case,
the term ${\mathcal{H}}_{\Delta }$ is inevitably present in the Hamiltonian,
and limits the total number of particles $N$. Namely, the condition ${%
\mathcal{H}}_{\Delta }<\Delta \varepsilon _{\mathrm{qh}}\lesssim \hbar
\omega _{c}$ and the fact that single-particle states with angular momenta
up to $L_{z}=3N(N-1)/2$ contribute to the states (\ref{psilaugh}) and (\ref%
{psiqh}), impose the constraint $3N(N-1)/2<\Delta \varepsilon _{\mathrm{qh}%
}/\Delta \omega $. For $\Delta \omega /\omega _{0}=10^{-3}$, it gives $N<30$.
Fortunately, this bound is large enough to expect the validity of our
calculations, which were performed for a homogeneous gas in the
thermodynamic limit.

In the experiments of the ENS group \cite{Dalibard:2004aa,Dalibard:2000aa},
the bosonic gas sample was brought into rotation by stirring it with an
additional laser. In addition to the harmonic potential of the optical trap,
there is an extra (quartic) confining potential that allows to reach and
even exceed the critical value $\omega _{c}$. In the case of critical
rotation, the term ${\mathcal{H}}_{\Delta }$ can be neglected and the number
of particles is only limited by the radial size of the gas cloud. We would
like to point out that in experiments of this type, it is possible to impose
a quenched disorder potential in the rotating frame, generated by speckle
radiation from a rotating diffractive mask \cite{speckle}. The rotation of
the mask should be synchronized with the stirring laser. This quenched
disorder potential localizes single-particle excitations that appear in the
system when the filling factor $\nu $ deviates from the value $1/3$.
Therefore, it provides fractional quantum Hall states with the robustness
necessary for experimental observation.

Let us finally discuss possible ways of experimental detection of the
fractional quantum Hall states. One of them could be the measurement of the
statistics of quasiholes using the Ramsey-type interferometric method
proposed in Ref. \cite{Cirac:2001aa}. Another possibility would be to study
the properties of the surface (edge) modes, which are analogous to the
chiral edge states of electrons in quantum Hall effect. The corresponding
analysis for a rotating bosonic cloud was recently performed in Ref. \cite%
{Cazalilla}. Finally, we could propose the detection of collective modes
that are similar to magnetorotons and magnetoplasmons collective modes in
electron quantum Hall systems (see, e.g., Girvin's contribution to Ref. \cite%
{Prange:1987aa}).

\begin{acknowledgments}
We are indebted to W. Apel for valuable advice and help, and we thank M.
Leduc, C. Salomon, L. Sanchez-Palencia, and G.V. Shlyapnikov
for helpful discussions. We
acknowledge support from the Deutsche Forschungsgemeinschaft SPP1116 and SFB
407, the RTN Cold Quantum Gases, ESF PESC BEC2000+, the Russian Foundation
for Basic Research, and the Alexander von Humboldt Foundation.
\end{acknowledgments}


\begin{thebibliography}{99}
\bibitem{Jaksch:1998aa} D. Jaksch, C. Bruder, J.I. Cirac, C.W. Gardiner, and
P. Zoller, Phys. Rev. Lett. \textbf{81},3108 (2000).

\bibitem{Greiner:2002aa} M. Greiner, O. Mandel, T. W. Ha"nsch, and I. Bloch,
Nature \textbf{415}, 39 (2002).

\bibitem{Dalibard:2004aa} V. Bretin, S. Stock, Y. Seurin, and J. Dalibard,
Phys. Rev. Lett. \textbf{92}, 050403 (2004).

\bibitem{Dalibard:2000aa} K.W. Madison, F. Chevy, W. Wohlleben, and J.
Dalibard, Phys. Rev. Lett. \textbf{84}, 806 (2000).

\bibitem{Cornell:2001aa} P.C. Haljan, I. Coddington, P. Engels, and E.A.
Cornell, Phys. Rev. Lett. \textbf{87}, 210403 (2000).

\bibitem{Cornell:2004aa} V. Schweikhard, I. Coddington, P. Engels, V.P.
Mogendorff, and E.A. Cornell, Phys. Rev. Lett. \textbf{92}, 040404 (2004).

\bibitem{Prange:1987aa} R.E. Prange and S.M. Girvin (editors), \textit{The
Quantum Hall Effect}, New York: Springer Verlag (1987).

\bibitem{Wilkin:2000aa} N.K. Wilkin and J.M.F. Gunn, Phys. Rev. Lett. 
\textbf{84}, 6 (2000).

\bibitem{Wilkin:2001aa} N.R. Cooper, N.K. Wilkin, and J.M.F. Gunn, Phys.
Rev. Lett. \textbf{87}, 120405 (2001).

\bibitem{Cirac:2001aa} B. Paredes, P. Fedichev, J.I. Cirac, and P. Zoller,
Phys. Rev. Lett. \textbf{87}, 010402 (2001).

\bibitem{Jain} J.K. Jain, Phys. Rev. Lett. \textbf{63}, 199 (1989).

\bibitem{Jolicoeur:2004aa} T. Jolicoeur and N. Regnault, e-print
cond-mat/0404093.

\bibitem{Bohn} C.A. Regal, C. Ticknor, J.L. Bohn, and D.S. Jin, Phys. Rev.
Lett. \textbf{90}, 053201 (2003).

\bibitem{nobel} M. Baranov, L. Dobrek, K. Goral, L. Santos, and M.
Lewenstein, Physica Scripta T \textbf{102}, 74 (2002).

\bibitem{Laughlin:1983aa} R.B. Laughlin, Phys. Rev. Lett. \textbf{50}, 1395
(1983).

\bibitem{Wigner} H. Fukuyama, Solid State Commun. \textbf{19}, 551 (1976);
M. Jonson and G. Srinivasan, Solid State Commun. \textbf{24}, 61 (1977).

\bibitem{lindemann} J.M. Ziman, \textit{Principles of the Theory of Solids},
Cambridge University Press, 1972.

\bibitem{Lozovik} Yu.E. Lozovik and V.M. Farztdinov, Solid State Commun. 
\textbf{54}, 725 (1985).

\bibitem{MonteCarlo} S.M. Girvin and T. Jach, Phys. Rev. B \textbf{29}, 5617
(1984).

\bibitem{Girvin:1984aa} S.M. Girvin, Phys. Rev. B \textbf{30}, 558 (1984).

\bibitem{speckle} P. Horak, J.-Y. Courtois, and G. Grynberg, Phys. Rev. A 
\textbf{58}, 3953 (1998); G. Grynberg, P. Horak, C. Mennerat-Robilliard,
Europhys. Lett. \textbf{49}, 424 (2000).

\bibitem{Cazalilla} M.A. Cazalilla, e-print cond-mat/0207715.
\end{thebibliography}
\end{document}